\documentclass[english,aps,floats,onecolumn,showpacs,nofootinbib]{revtex4}
\usepackage{pslatex}
\usepackage[T1]{fontenc}
\usepackage[latin1]{inputenc}
\usepackage{graphicx}
\usepackage{epsfig}

\usepackage{calc}
\usepackage{ifthen}

{
{
{
\newcommand{\bea}{\begin{eqnarray}}
\newcommand{\eea}{\end{eqnarray}}

\newcommand{\nc}{\newcommand}
\nc{\renc}{\renewcommand}
\nc{\eqs}[2]{\mbox{Eqs.~(\ref{#1},\,\ref{#2})}}
\nc{\eq}[1]{\mbox{Eq.~(\ref{#1})}}
\nc{\figs}[2]{\mbox{Figs.~(\ref{#1},\,\ref{#2})}}
\nc{\fig}[1]{\mbox{Fig~.(\ref{#1})}}
\nc{\be}[1]{\begin{equation} \mbox{$\label{#1}$}}
\nc{\ee}{\vspace{0.1cm}\end{equation}}

\newcommand{\bean}{\begin{eqnarray*}}
\newcommand{\eean}{\end{eqnarray*}}

\def\eV{{\rm \ eV}}
\def\GeV{{\rm \ GeV}}

\def\keV{{\rm \ keV}}
\def\TeV{{\rm \ TeV}}

\def\lae{\;^{<}_{\sim} \;} \def\gae{\; ^{>}_{\sim} \;}

\begin{document}
\title{Warm Dark Matter via Ultra-Violet Freeze-In: Reheating Temperature and Non-Thermal Distribution for Fermionic Higgs Portal Dark Matter}
\author{John McDonald}
\email{j.mcdonald@lancaster.ac.uk}
\affiliation{Dept. of Physics, University of 
Lancaster, Lancaster LA1 4YB, UK}

\begin{abstract}

    Warm dark matter (WDM) of order keV mass may be able to resolve the disagreement between structure formation in cold dark matter simulations and observations. The detailed properties of WDM will depend upon its energy distribution, in particular how it deviates from the thermal distribution usually assumed in WDM simulations. Here we focus on WDM production via the Ultra-Violet (UV) freeze-in mechanism, for the case of fermionic Higgs portal dark matter $\psi$ produced via the portal interaction $\overline{\psi}\psi H^{\dagger}H/\Lambda$. We introduce a new method to simplify the computation of the non-thermal energy distribution of dark matter from freeze-in. We show that the non-thermal energy distribution from UV freeze-in is hotter than the corresponding thermal distribution and has the form of a Bose-Einstein distribution with a non-thermal normalization.  The resulting range of dark matter fermion mass consistent with observations is 5-7 keV. The reheating temperature must satisfy $T_{R} \gae 120 \GeV$ in order to account for the observed dark matter density when $m_{\psi} \approx 5 $ keV, where the lower bound on $T_{R}$ corresponds to the limit where the fermion mass is entirely due to electroweak symmetry breaking via the portal interaction. The corresponding bound on the interaction scale is $\Lambda \gae 6.0 \times 10^{9} \GeV$.

\end{abstract}
 
\maketitle

\section{Introduction}

   Cold dark matter (CDM) is successful in explaining structure on large cosmological scales, but N-body CDM simulations do not reproduce the cored nature of observed galaxies \cite{salucci} and also produce significantly more satellite-sized subhalos in Milky Way-sized galaxies than the number of observed satellites of the Milky Way \cite{boylan}. Warm dark matter (WDM) \cite{wdm1, wdm2} has been proposed as a solution to the problems of CDM. Dark matter particles with order keV mass can reduce structure on small scales by free-streaming, in principle suppressing the formation of subhalos and of cusps at halo centres. The free-streaming of WDM can be studied by observation of the Lyman-$\alpha$ forest, which results in a lower bound on the WDM mass of 2-3 keV \cite{seljak, riotto1, viel}. (See also \cite{sanchez1,sanchez2}.) Bounds can also be obtained by using the statistics of primeval galaxies at high redshift combined with constraints from re-ionisation on the number density of small halo galaxies at $z \gae 4$ \cite{pacucci,schultz}. In \cite{lapi}, a lower bound on the WDM particle mass $m_{WDM} \gae 2$ keV was obtained. An upper bound $m_{WDM} \lae 3$ keV is obtained from the Planck 1-$\sigma$ upper limit on the electron scattering optical depth and the requirement that WDM can suppress the number of satellites in Milky Way-sized galaxies \cite{lapi}. Thus the range of mass for which WDM is consistent with observations is 2-3 keV, assuming a thermal energy distribution for the dark matter \footnote{While WDM in this range can suppress large subhalo formation, its free-streaming length is too small to solve the cusp-core problem \cite{brooks}. In \cite{sanchez3} it is argued that quantum pressure can suppress the formation of cusps in the case of fermionic WDM.} \cite{lapi}.

   Therefore there is good reason to investigate dark matter production mechanisms that can generate keV scale dark matter.  Examples include non-resonant \cite{dw} and resonant \cite{sf} production of sterile neutrino dark matter via oscillations, sterile neutrino production via inflaton decay \cite{shap}, decay of singlet scalars to sterile neutrinos \cite{kuz,kuz2,merle}, decay of thermal plasma scalars to sterile neutrinos \cite{drewes}, and a two-step model where singlet scalars are produced via freeze-in and subsequently decay to sterile neutrinos \cite{merle2}. An important question is the energy distribution of the WDM. Structure formation is sensitive to the momentum distribution of the WDM particles. The usual assumption in simulations is that WDM is a fermion with a thermal distribution. Thus it is important to determine whether the energy distribution for a given WDM production mechanism deviates from a thermal distribution. In this paper we will investigate a specific production mechanism, Ultra-Violet (UV) freeze-in, and a specific WDM candidate, fermionic Higgs portal dark matter. 

The now standard freeze-in mechanism for the case of renormalizable interactions, where a non-equilibrium density of dark matter particles is accumulated via annihilations or decays of thermal background particles,  was first applied to the specific case of the renormalizable Higgs portal in \cite{jm1} and later generalized in \cite{fi1}, where it was first called "freeze-in". The freeze-in mechanism for non-renormalizable interactions was first applied to fermionic dark matter in \cite{jn1}, in the context of SUSY Higgs portal dark matter. It was later discussed in \cite{fi1} and a more general analysis of the mechanism was given in \cite{uvfi}, where it was called the "Ultra-Violet (UV) freeze-in" mechanism, a term which we will use in the following. Essentially the same mechanism was applied much earlier to the case of thermal gravitino production \cite{ellisg}.

     The UV freeze-in mechanism produces particles mostly during the transition from inflaton domination to radiation domination, at around the reheating temperature $T_{R}$. Since the dark matter is never in thermal equilibrium, the resulting energy distribution will be non-thermal. In the following we will determine the dark matter energy distribution produced by the UV freeze-in mechanism and the constraints on the reheating temperature necessary to produce keV scale fermionic Higgs portal dark matter. 

The paper is organized as follows. In Section 2 we use a threshold approximation to analytically determine the dark matter density from UV freeze-in and to obtain the reheating temperature $T_{R}$ and the non-renormalizable portal interaction scale $\Lambda$ necessary to account for the observed dark matter density. In Section 3 we introduce a new method to simplify the computation of the non-thermal energy distribution of dark matter from UV freeze-in, which we apply to determine the energy distribution for fermionic Higgs portal dark matter.  In Section 4 we present our conclusions.

\section{Fermionic Higgs portal Dark matter production via UV freeze-in} 

The fermionic Higgs portal model is defined by the following interaction and mass terms
\be{d1} {\cal L} \supset \frac{1}{\Lambda} \overline{\psi}{\psi} H^{\dagger}H  + m_{\psi_{0}} \overline{\psi} \psi   ~.\ee
In the following we will consider $\psi$ to be a Dirac fermion (the analysis is similar for a Majorana fermion). After electroweak symmetry breaking, the $\psi$ mass is given by 
\be{d2} m_{\psi} = m_{\psi_{0}} + m_{\psi_{sb}} =  m_{\psi_{0}} + \frac{v^2}{2 \Lambda}   ~,\ee
where $<H> = v/\sqrt{2}$, $v = 246 \GeV$ and $m_{\psi_{sb}}$ is the $\psi$ mass purely from electroweak symmetry breaking.  Since we will be considering annihilation at temperatures large compared to $v$,  we can consider $H$ to be two equivalent complex scalars, $\phi_{i}$ ($i = 1,\;2$). The relativistic cross-section for annihilations $\phi_{i}^{\dagger}\phi_{i} \rightarrow \overline{\psi} \psi$  for each complex scalar is\footnote{In this we have used the zero-temperature annihilation cross-section. As shown in \cite{drewes}, the main effect of finite-temperature corrections, in the case where the final state density is much below the equilibrium density, is the finite-temperature quasiparticle scalar mass. Since we are considering relativistic $\phi$ scalars with mean energy $\sim 3 T$, we do not expect the finite-temperature quasiparticle mass ($M_{\phi}^{2} = \lambda T^{2}/24$ for a $\lambda \phi^{4}/4!$ interaction) to have a significant effect on the annihilation rate.}  
\be{d3} \sigma_{\phi} = \frac{1}{8 \pi \Lambda^2}    ~.\ee 
Therefore the total production rate of $\psi$ fermions due to both complex scalars is described by the rate equation
\be{d4} \frac{d n_{\psi}}{d t} + 3 H n_{\psi} = 2  <\sigma_{\phi} v_{rel}> n_{\phi_{eq}}^{2} \equiv \frac{n_{\phi_{eq}}^2 }{4 \pi \Lambda^2} ~,\ee  
where $n_{\phi_{eq}} = \zeta(3) T^{3}/\pi^2$ ($\zeta(3) \approx 1.2$) is the equilibrium number density of each scalar $\phi_{i}$ and we have used $<v_{rel}> = 1$ for highly relativistic annihilations \cite{cannoni}.  In terms of the scale factor, and using \eq{d3}, \eq{d4} becomes 
\be{d5} \frac{H}{a^2} \frac{d(n_{\psi} a^3)}{da} = \frac{1}{4 \pi \Lambda^2} \left( \frac{1.2 T^3}{\pi^2} \right)^{2}   ~.\ee 
 The reheating temperature, $T_{R}$, is defined by the inflaton decay rate $\Gamma_{d}$ via $H(T_{R}) = \Gamma_{d}$, where $H(T_{R})$ is calculated for a radiation-dominated energy density\footnote{We can equivalently consider $T_{R}$ to define the inflaton decay rate via $\Gamma_{d} = H(T_{R})$.}. (We consider $\Gamma_{d}$ to be a constant in the following.) Therefore  
\be{d6} T_{R} = \left(\frac{M_{P} \Gamma_{d}}{k_{T_{R}}} \right)^{1/2}  ~,\ee
where $M_{P} = (8 \pi G)^{-1/2}$, $k_{T} = \left(\pi^2 g(T)/90 \right)^{1/2}$ and  $g(T)$ is the number of relativistic degrees of freedom in the thermal bath. (We will assume that $g(T) = g(T_{R})$ throughout the freeze-in process.) We next solve \eq{d5} by using a threshold approximation, where  the Universe is considered to be entirely radiation-dominated for $T < T_{R}$ and entirely inflaton-dominated for $T > T_{R}$. 
 (We will later show that this is in reasonable agreement with the exact numerical solution.)

During the radiation-dominated era at $T < T_{R}$ we have $H = k_{T_{R}} T^2/M_{p}$ and $a \propto T^{-1}$. \eq{d5} then becomes 
\be{d7} \frac{d}{dT} \left( \frac{n_{\psi}}{T^3} \right) = -  \frac{1}{4 \pi \Lambda^2} \left( \frac{1.2}{\pi^2}\right)^{2} 
\frac{T^2}{H} \equiv    -  \frac{1}{4 \pi \Lambda^2} \left( \frac{1.2}{\pi^2}\right)^{2} 
\frac{M_{P}}{k_{T_{R}}}   ~.\ee
Therefore
\be{d8} \frac{n_{\psi}}{T^3} = \left(\frac{n_{\psi}}{T^
3}\right)_{T_{R}} +  \frac{1}{4 \pi \Lambda^2} 
\left( \frac{1.2}{\pi^2}\right)^{2} 
\frac{M_{P}}{k_{T_{R}}} \left(T_{R} - T \right)  ~.\ee
Therefore at  $T \ll T_{R}$, when the comoving density $a^3 \times n_{\psi}$ has become constant, the number density is given by
\be{d9} \frac{n_{\psi}}{T^3} = \left(\frac{n_{\psi}}{T^3}\right)_{T_{R}} +  \frac{1}{4 \pi \Lambda^2} \left( \frac{1.2}{\pi^2}\right)^{2} 
\frac{M_{P} T_{R}}{k_{T_{R}}}  ~.\ee
We next determine $(n_{\psi}/T^3)_{T_{R}}$. During the inflaton-dominated era at $T > T_{R}$, the radiation background due to inflaton decays is described by $a \propto T^{-8/3}$, where $T = k_{r} \left(M_{P} H T_{R}^{2}\right)^{1/4}$ and $k_{r} = (9/5 \pi^3 g(T_{R}))^{1/8}$. In this case \eq{d5} becomes 
\be{d10} \frac{d}{dT} \left(\frac{n_{\psi}}{T^8}\right) = -\frac{8}{3 H T^9}\frac{1}{4 \pi \Lambda^2} \left( \frac{1.2 T^3}{\pi^2} \right)^{2}  \equiv - \frac{2 k_{r}^{4} M_{P} T_{R}^2}{3 \pi \Lambda^2}\left( \frac{1.2}{\pi^2} \right)^{2}   \frac{1}{T^7} \equiv - \frac{\alpha_{T}}{T^7}    ~,\ee
where we have defined $\alpha_{T}$ for convenience.  
Therefore, on integrating from $T_{i}$ to $T_{R}$ we obtain  
\be{d11} \left(\frac{n_{\psi}}{T^8} \right)_{T_{R}} =  \left(\frac{n_{\psi}}{T^8} \right)_{T_{i}} + 
\frac{\alpha_{T}}{6}\left(\frac{1}{T_{R}^{6}} - \frac{1}{T_{i}^{6}} \right)   ~.\ee
Assuming that initially $T_{i} \gg T_{R}$ and that 
$n_{\psi}/T^8$ at $T_{i}$ is negligible, 
we obtain
\be{d12} \left(\frac{n_{\psi}}{T^3} \right)_{T_{R}} = 
 \frac{\alpha_{T}}{6 T_{R}} =  
\frac{1}{4 \pi \Lambda^2}\left( \frac{1.2}{\pi^2} \right)^{2} 
\frac{M_{P} T_{R}}{k_{T_{R}}} \times \frac{4}{9} \,k_{r}^4 k_{T_{R}}    ~.\ee
Thus the total $\psi$ density from UV freeze-in is 
\be{d13} \frac{n_{\psi}}{T^3} = 
\frac{1}{4 \pi \Lambda^2 }\left( \frac{1.2}{\pi^2} \right)^{2} 
\frac{M_{P} T_{R}}{k_{T_{R}}} \left(1 + \frac{4}{9}\, k_{r}^4 k_{T_{R}} \right)    ~.\ee 
Since $k_{r}^{4}k_{T_{R}} = (1/50 \pi)^{1/2}$, we obtain
\be{d14} \frac{n_{\psi}}{T^3} = (1 + 0.035)\frac{1}{4 \pi \Lambda^2 }\left( \frac{1.2}{\pi^2} \right)^{2} 
\frac{M_{P} T_{R}}{k_{T_{R}}} ~,\ee
where 0.035 is the contribution due to $\psi$ production during inflaton-domination. Thus almost all of the dark matter production occurs during radiation-domination at $T < T_{R}$. 

In general, we can parameterize the dark matter density from UV freeze-in by a factor $\gamma_{R}$, such that 
\be{d15} \frac{n_{\psi}}{T^3} = \frac{\gamma_{R}}{4 \pi \Lambda^2 }\left( \frac{1.2}{\pi^2} \right)^{2} 
\frac{M_{P} T_{R}}{k_{T_{R}}} ~,\ee
where $\gamma_{R} = 1.035$ for the threshold approximation. We will show later that that $\gamma_{R} = 0.73$ from the exact numerical solution of \eq{d5}. Therefore the threshold approximation is in reasonable agreement with the numerical solution and provides a useful analytical expression for the dark matter density.

We next determine the constraints on $\Lambda$ and $T_{R}$ necessary to account for keV mass dark matter. The total fermion dark matter density $\Omega_{\psi}$ is due to $\psi$ plus $\overline{\psi}$. Therefore at the present CMB temperature, $T_{\gamma}$,  
\be{d16} \Omega_{\psi} = \frac{ 2 m_{\psi} n_{\psi}(T_{\gamma})}{\rho_{c}}  ~,\ee
where $\rho_{c} = 8.1\times 10^{-47} h^2  \; \GeV^4$ is the critical density and $H_{0} = 100 h \; {\rm km} {\rm s^{-1}} {\rm Mpc^{-1}}$. $n_{\psi}(T_{\gamma})$ is given by   
\be{d17} \frac{n_{\psi}(T_{\gamma})}{T_{\gamma}^{3}}   
= \frac{g(T_{\gamma})}{g(T_{R})} \frac{n_{\psi}(T)}{T^{3}} =  
\frac{g(T_{\gamma})}{g(T_{R})}\frac{\gamma_{R}}{k_{T_{R}}}\left( \frac{1.2}{\pi^2} \right)^{2} 
\frac{M_{P} T_{R}}{4 \pi \Lambda^2}  ~.\ee
Here $T$ is chosen such that the comoving number density $n_{\psi}a^3$ is frozen but $T$ is sufficiently close to $T_{R}$ that $g(T) = g(T_{R})$.
Therefore 
\be{d18} \Omega_{\psi} = \left(\frac{m_{\psi_{sb}}}{m_{\psi}} \right)^{2} \frac{g(T_{\gamma})}{g(T_{R})} \left( \frac{1.2}{\pi^2} \right)^{2} \frac{2 \gamma_{R} T_{\gamma}^{3} T_{R} M_{P} }{\pi k_{T_{R}} \rho_{c} v^4} \times m_{\psi}^3   ~. \ee
Thus requiring that $\Omega_{\psi} = \Omega_{DM \;obs}$ gives for the reheating temperature 
\be{d19} 
T_{R} = \left(\frac{m_{\psi}}{m_{\psi_{sb}}} \right)^{2} \frac{g(T_{R})}{g(T_{\gamma})}\left( \frac{\pi^2}{1.2} \right)^{2} \frac{\pi k_{T_{R}} \rho_{c} v^4}{2 \gamma_{R} T_{\gamma}^{3} M_{P}}    \frac{\Omega_{DM \; obs}}{m_{\psi}^3}  ~.\ee
As input we use $h = 0.68$, $\Omega_{DM \; obs} = 0.27$ and $g(T_{R}) = 106.75$, which gives
\be{d20} T_{R} = \frac{1.38}{\gamma_{R}} \left( \frac{2 \keV}{m_{\psi}}\right)^{3}  
\left(\frac{m_{\psi}
}{m_{\psi_{sb}}} \right)^{2} \left( \frac{\Omega_{DM\;obs}}{0.27} \right)\; \TeV  ~,\ee
or, equivalently, using $m_{\psi_{sb}} = v^{2}/2 \Lambda$,
\be{d20a} T_{R} = \frac{1.38}{\gamma_{R}}\left( \frac{2 \keV}{m_{\psi}}\right) \left(\frac{\Lambda}{1.5 \times 10^{10} \GeV} \right)^{2}
\left( \frac{\Omega_{DM\;obs}}{0.27} \right)\; \TeV  ~.\ee
We see that the ratio $T_{R}/\Lambda^2$ is fixed for a given $m_{\psi}$. Since in general we have $m_{\psi} \geq m_{\psi_{sb}}$, it follows that $T_{R} \gae 1.3 \TeV$ (using $\gamma_{R} = 1.035$), where the lower bound is the limit where $\psi$ gains its mass entirely through electroweak symmetry breaking via the portal interaction. 

The portal interaction mass scale $\Lambda$ is given by 
\be{d21} \Lambda = \frac{v^2}{2 m_{\psi}} \left(\frac{m_{\psi}}{m_{\psi_{sb}}}\right)  = 1.5 \times 10^{10} \left(\frac{2 \keV}{m_{\psi}}\right)\left(\frac{m_{\psi}}{m_{\psi_{sb}}}\right) \GeV   ~.\ee
Thus there is a corresponding lower bound on the interaction mass scale for keV mass dark matter, $\Lambda \gae 10^{10}$ GeV. 

The case where the fermion gains its mass entirely from electoweak symmetry breaking is particularly interesting, as one way to understand for the lightness of the dark matter fermion is via the smallness of the electroweak scale relative to a large mass scale $\Lambda$ for the portal interaction. In this case a low reheating temperature $T_{R} \sim 1.3 \TeV$ is predicted.

  The above analysis assumes that $\psi$ never reaches its thermal equilibrium density.  This requires that 
$n_{\psi} < n_{\psi_{eq}}$ where $n_{\psi_{eq}} = 3\zeta(3)T^3/2 \pi^2$. Using \eq{d15}, this condition becomes 
\be{d22} T_{R} < \frac{6}{\zeta(3)} \frac{\pi^3 k_{T_{R}} \Lambda^2}{\gamma_{R} M_{P}}   ~.\ee 
Using \eq{d20a}, we find that this is satisfied if 
\be{d23} m_{\psi}  > 56 \left( \frac{\Omega_{DM\;obs}}{0.27} \right) \; \eV   ~.\ee
This is easily satisfied when $m_{\psi} \approx 2$ keV.

\section{Non-thermal energy distribution from Ultra-Violet Freeze-In} 

     In order to calculate the non-thermal energy distribution of the $\psi$ particles from $\phi_{i}$ annihilation, we use the following method based on the rate equation. The Higgs scalars $\phi_{i}$ have rapid interactions with the Standard Model particles in the thermal bath, with the time scale of the interactions given dimensionally by $T^{-1}$. The time scale of the annihilation process to $\overline{\psi}\psi$ pairs is much longer (since the $\psi$ particles are out-of-equilibrium). We therefore first consider the production of $\psi$ pairs on a time scale $\Delta t$ which is long compared to $T^{-1}$ but short compared to $H^{-1}$. During $\Delta t$ we can therefore neglect the expansion of the Universe, while the number density of $\phi$ particles will be maintained at its equilibrium value by its rapid interactions with the thermal bath. Once the number density of $\psi$ particles produced during $\Delta t$ is known, the effect of expansion can then be included by integrating over the production time steps $\Delta t$ and including the scaling of the produced $\psi$ density and the background temperature with respect to scale factor, in order to obtain the total $\psi$ density from UV freeze-in.

    The production of $\psi$ dark matter occurs mostly close to the reheating temperature. Its total rate of production by  identical annihilating particles in thermal equilibrium $\phi_{i}$, $\sum_{i} \phi_{i}^{\dagger}\phi_{i} \rightarrow \overline{\psi}\psi$, satisfies the equation 
\be{n1} \frac{dn_{\psi}}{dt} = - \sum_{i} \frac{dn_{\phi_{i}}}{dt} \equiv N <\sigma_{\phi} v_{rel}> n_{\phi_{eq}}^{2}    ~,\ee
where $n_{\phi_{eq}}$ is the equilibrium number density of each identical particle $\phi_{i}$ and $N$ is the number of such particles  ($i = 1,...,N$). This is the rate equation for the production of $\overline{\psi}\psi$ pairs for the case where the time scale $\Delta t$ during which particles are produced is short enough that expansion may be neglected.   

We are interested in the energy distribution $f_{\psi}(E)$ of relativistic $\psi$ particles produced by UV freeze-in. In general, for effectively massless relativistic particles,  
\be{n2} n_{\psi} = \int_{0}^{\infty} f_{\psi}(E) dE    ~,\ee
where $f_{\psi}(E) = dn_{\psi}/dE$.  We first introduce a   method to simplify the computation of $f_{\psi}(E)$. We assume throughout that $<\sigma_{\phi} v_{rel}>$ is independent of energy. We first re-write \eq{n1} as 
\be{n2a}  \int_{0}^{\infty} \frac{df_{\psi}}{dt} dE = N <\sigma_{\phi} v_{rel}>  n_{\phi_{eq}} \int_{0}^{\infty} f_{\phi_{eq}} dE   ~.\ee 
Therefore    
\be{n3} \int_{0}^{\infty} \left[ \frac{df_{\psi}}{dt} - 
N n_{\phi_{eq}} <\sigma_{\phi} v_{rel}>  f_{\phi_{eq}}  \right] dE  = 0   ~.\ee
Thus, in general
\be{n4}   \frac{df_{\psi}}{dt} - 
 N n_{\phi_{eq}} <\sigma_{\phi} v_{rel}>  f_{\phi_{eq}}  = g(E)  ~,\ee
where $g(E)$ is a function such that
\be{n5} \int_{0}^{\infty} g(E) dE  = 0  ~.\ee
We next show that $g(E) = 0$. The annihilation process implies that the reduction in the energy density of the $\phi_{i}$ particles is equal to the increase in the energy density of the $\psi$ particles. In general, the energy density in the $\phi$ particles is 
\be{n4a} \rho_{\phi} = N \int_{0}^{\infty} f_{\phi} E dE = N \overline{E} \times n_{\phi_{i}}   ~,\ee
where $\overline{E}$ is the mean energy per $\phi_{i}$ particle
\be{n5b} \overline{E} = \frac{1}{n_{\phi_{i}}} \int_{0}^{\infty} f_{\phi} E dE     ~.\ee
Since the $\phi_{i}$ particles are in thermal equilibrium, and we are considering production of $\psi$ particles via annihilation on a time scale large compared to $T^{-1}$, the mean energy $\overline{E}$ of the $\phi_{i}$ particles will be kept constant throughout. Therefore
\be{n5a} \frac{d \rho_{\psi}}{dt} = -\frac{d \rho_{\phi_{i}}}{dt} =  -N \overline{E} \times \frac{d n_{\phi_{i}}}{dt}    ~,\ee
where $d n_{\phi_{i}}/dt$ represents the rate of annihilation of $\phi_{i}$ particles to $\overline{\psi} \psi$ pairs, 
\be{n4b} \frac{dn_{\phi_{i}}}{dt} = -n_{\phi_{eq}}^{2} <\sigma_{\phi} v_{rel}> ~.\ee
Using \eq{n1} and   \eq{n4b}, \eq{n5a} becomes
\be{n5b} \frac{d \rho_{\psi}}{dt} \equiv \frac{d}{dt}\int_{0}^{\infty} f_{\psi}(E) E dE 
= \frac{N}{n_{\phi_{eq}}} \int_{0}^{\infty} f_{\phi_{eq}} E dE \times n_{\phi_{eq}}^2 <\sigma_{\phi} v_{rel} >    ~.\ee 
Therefore 
\be{n5c} \int_{0}^{\infty} \frac{d f_{\psi}(E)}{dt}  E dE 
= \int_{0}^{\infty} N n_{\phi_{eq}} <\sigma_{\phi} v_{rel} > f_{\phi_{ eq}}\, EdE   ~.\ee
Comparing with \eq{n4}, this implies that 
\be{n5d}  \int_{0}^{\infty} g(E)E dE  = 0  ~.\ee
In order for the integral of $g(E)$ to equal zero when $g(E) \neq 0$, the integral much consist of positive and negative contributions which exactly cancel out. However, in this case the integral of $g(E)E$ will not cancel out. Therefore to simultaneously satisfy \eq{n5} and \eq{n5d} it is necessary that $g(E) = 0$. Thus we obtain our main result for $\overline{\psi} \psi$ production during an interval $\Delta t$ such that $T^{-1} \ll \Delta t  \ll  H^{-1}$, 
 \be{n6}   \frac{df_{\psi}}{dt} = 
 N n_{\phi_{eq}} <\sigma_{\phi} v_{rel}>  f_{\phi_{eq}} ~.\ee

We next include expansion and calculate the total freeze-in $\psi$ density. During an increment of scale factor $\Delta a$, \eq{n6} can be written as 
\be{n9}  \frac{df_{\psi}}{da} =
 \frac{N n_{\phi_{eq}} <\sigma_{\phi} v_{rel}> }{aH}   f_{\phi_{eq}}   ~.\ee
Thus 
\be{n10} \Delta f_{\psi}(E(a)) = \frac{ N n_{\phi_{eq}} <\sigma_{\phi} v_{rel}> }{aH(a)} f_{\phi_{eq}}(E(a)) \Delta a  ~.\ee
Writing $f_{\psi} = dn_{\psi}/dE$, we can say that  
\be{n11}  \Delta n_{\psi}(E(a))  = \frac{ N n_{\phi_{eq}} <\sigma_{\phi} v_{rel}> }{aH(a)} f_{\phi_{eq}}(E(a))\Delta E  \Delta a  ~,\ee
where $\Delta n(E(a))$ is the contribution to the $\psi$ number density in the range $\Delta E$ around $E(a)$ produced during $\Delta a$ at $a$. This then is diluted by subsequent expansion, while the energy decreases by redshifting. Therefore at a later time with scale factor $a_{0} < a$, we have 
$\Delta n_{\psi}(a_{0}) = (a/a_{0})^{3} \Delta n_{\psi}(a)$, $E_{0} = (a/a_{0}) E$ and $\Delta E_{0} = (a/a_{0}) \Delta E$. Thus 
\be{n12}  \Delta n_{\psi}(a_{0})  = \left(\frac{a}{a_{0}}\right)^3 \frac{ N n_{\phi_{eq}} <\sigma_{\phi} v_{rel}> }{aH(a)} f_{\phi_{eq}}(E(a)) \left(\frac{a{_0}}{a}\right) \Delta E_{0}  \Delta a  ~.\ee
Therefore 
\be{n13}  \frac{\Delta n_{\psi}(a_{0})}{\Delta E_{0}}   = \left(\frac{a}{a_{0}}\right)^2\frac{ N n_{\phi_{eq}} <\sigma_{\phi} v_{rel}> }{aH(a)} f_{\phi_{eq}}(E(a)) \Delta a  ~.\ee
This is the contribution to $dn_{\psi}/dE_{0}$ from $\Delta a$ around $a$. To get the total energy distribution we integrate from an early scale factor $a_{i}$ during the inflation-dominated era to the late scale factor $a_{0}$, 
\be{n14}  \frac{dn_{\psi}}{d E_{0}}(a_{0}, E_{0}) = \int_{a_{i}}^{a_{0}} \left(\frac{a}{a_{0}}\right)^{2}
\left( \frac{N n_{\phi_{eq}} <\sigma_{\phi} v_{rel}>  f_{\phi_{eq}}}{aH}\right) da  ~.\ee

 We next apply this method to the case of thermal equilibrium Higgs scalars to annihilating to portal fermions. In this case $N = 2$.  To evaluate \eq{n14}, we need to integrate through the transition from inflaton to radiation-domination, during which most of the dark matter is produced. We therefore solve the coupled equations for the inflaton and radiation densities to obtain $T$ and $H$ as a function of $a$  
\be{n15} aH \frac{d \rho_{rad}}{da} + 4 H \rho_{rad} = \Gamma_{d} \rho_{inf}    ~\ee 
and  
\be{n16} aH \frac{d \rho_{inf}}{da} + 3 H \rho_{inf} = - \Gamma_{d} \rho_{inf}    ~,\ee  
where $H(a) = (\rho(a)/3 M_{P})^{1/2}$ and $\rho(a) = \rho_{inf}(a) + \rho_{rad}(a)$. The temperature is obtained from the radiation density via 
\be{n17} T(a) = \left(\frac{30}{\pi^2 g(T_{R})}\right)^{1/4} \rho_{rad}^{1/4}(a)   ~.\ee 
The equilibrium distribution for the scalars is 
\be{n18} f_{\phi_{eq}} = \frac{1}{2 \pi^2} \frac{E^2}{\left(e^{E/T} - 1\right)}    ~.\ee
Thus 
\be{n19}  \frac{dn_{\psi}}{d E_{0}}(a_{0}, E_{0}) = 
\frac{1.2}{8 \pi^5 \Lambda^2} 
\int_{a_{i}}^{a_{0}} \left(\frac{a}{a_{0}}\right)^2
\frac{T(a)^{3} E(a)^{2}}{\left( \exp\left(\frac{E(a)}{T(a)}\right) -1 \right) }
\frac{da}{a H(a)}  ~,\ee
where $E(a) = E_{0} (a_{0}/a)$. By numerically integrating \eq{n19} together with \eq{n15} and \eq{n16} we can obtain $dn_{\psi}/dE_{0}$ as a function of $E_{0}$. This gives the energy distribution of relativistic $\psi$ dark matter from UV freeze-in at a scale factor $a_{0}$ at which the comoving dark matter density $a^{3} n_{\psi}$ is frozen.

  To show the non-thermal nature of the resulting energy distribution, we compare it to the thermal $\psi$ distribution which gives the same total number density at $a_{0}$. (This corresponds to the case of thermal relic fermion WDM which decouples at a high temperature \cite{riotto1}.) To do this, we first integrate the $\psi$ 
energy distribution over $E_{0}$ to obtain the total number density and then set this equal to the equilibrium number density for relativistic fermions to determine an equivalent temperature $T_{eff}$ which produces the same number density 
\be{n20} n_{\psi\;eq} \equiv \frac{3}{4} \left(\frac{2 \zeta(3)}{\pi^2}\right) T_{eff}^3 = n_{\psi}(a_{0}) ~.\ee 
The equivalent thermal distribution is then the Fermi-Dirac distribution at $T_{eff}$ 
\be{n21} \left( \frac{dn_{\psi}}{d E_{0}} \right)_{thermal}   = \frac{1}{\pi^2} \frac{E_{0}^{2}}{ \left( \exp\left(\frac{E_{0}}{T_{eff}}\right) + 1 \right) }  ~.\ee

  In Figure 1 we show the non-thermal and equivalent thermal  distributions for the case of fermionic Higgs portal dark matter with $m_{\psi} = 2$ keV when $T_{0} = 20$ GeV and $g(T_{0}) = g(T_{R}) = 10^2$. (The distributions at other scale factors are obtained by simply scaling the axes as $dn/dE_{0} \propto a^{-2}$ and $E_{0} \propto a^{-1}$.) 
Both the non-thermal distribution and equivalent thermal distribution depend only on the ratio $T_{R}/\Lambda^2$, which is fixed by $m_{\psi}$ via \eq{d20a}. Therefore, at a fixed $T_{0}$, the distributions are independent of $T_{R}$ when dark matter is explained by portal fermions with $m_{\psi} = 2$ keV.  From Figure 1 we see that the distribution from UV freeze-in is much broader and shallower than the corresponding thermal distribution, corresponding to a hotter WDM distribution. The mean energy of the Higgs portal WDM is 
$\overline{E} =  2.34 T_{0}$, compared to the thermal distribution for which $\overline{E} = 0.95 T_{0}$, where we have used $T_{eff} = 0.03 T_{0}$, which we derive below.

\begin{figure}[htbp]
\begin{center}
\epsfig{file=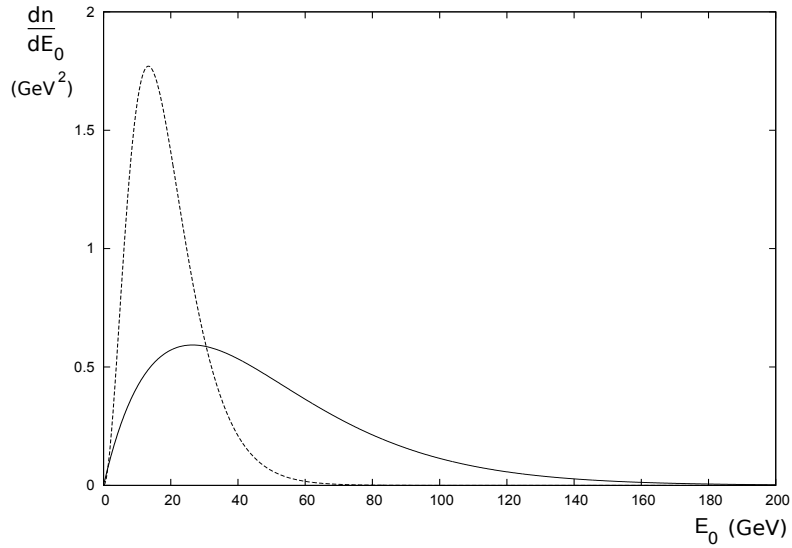, width=0.4\textwidth, angle = -90}
\caption{Non-thermal distribution from UV freeze-in (solid line) and corresponding Fermi-Dirac thermal distribution (dashed line) for the case of fermionic Higgs portal dark matter with $m_{\psi} = 2$ keV. The distributions are shown at $T_{0} = 20$ GeV.}
\label{fig1}
\end{center}
\end{figure}

In addition, we can compare the total number density  with that expected from the threshold approximation.  We find that $\gamma_{R} = 0.73$ for $m_{\psi} = 2$ keV dark matter. This in reasonable agreement with the threshold approximation value, $\gamma_{R} = 1.035$. With this value of $\gamma_R$, the lower bound on the reheating temperature becomes $T_{R} \gae 1.9 \TeV$.

We find that the distribution from UV freeze-in can be well-approximated by
\be{n22} \frac{dn}{dE_{0}} \approx \frac{1.29\,T_{R}M_{P}}{8 \pi^5 \Lambda^2 k_{T_{R}} } \times \frac{E_{0}^2}{\left(\exp\left(\frac{1.155\,E_{0}}{T_{0}}\right) - 1\right) }    ~.\ee
This depends only on the ratio $T_{R}/\Lambda^2$. The distribution has the form of a Bose-Einstein distribution with non-thermal normalization and modified temperature. In Figure 2 we show the exact numerical solution and the approximate solution.

\begin{figure}[htbp]
\begin{center}
\epsfig{file=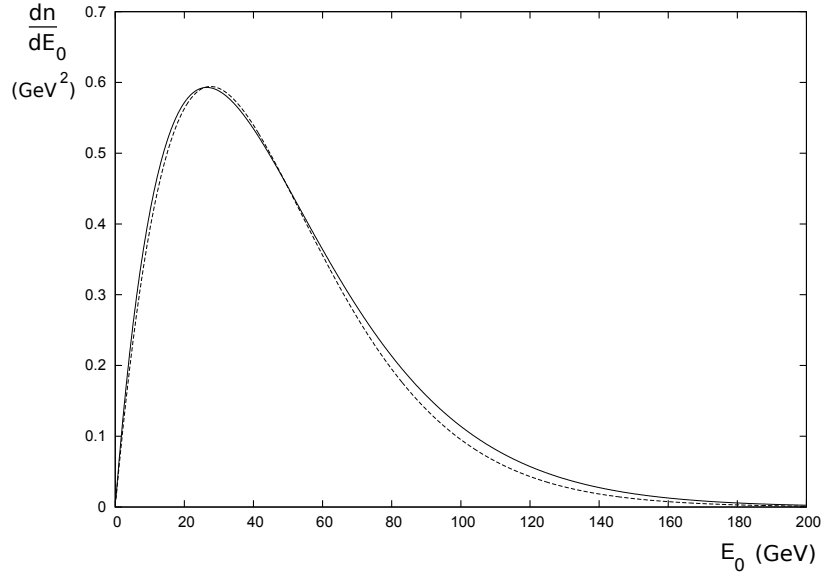, width=0.6\textwidth, angle = 0}
\caption{Comparison of the exact numerical (solid line) and approximate analytical solution (dashed line) for the non-thermal distribution from UV freeze-in. ($m_{\psi} = 2$ keV, $T_{0} = 20$ GeV.)}
\label{fig2}
\end{center}
\end{figure}

We can also use the threshold approximation to find an analytical expression for $T_{eff}$.  Using \eq{d15} for $n_{\psi}(a_{0})$ in \eq{n20}, we obtain 
\be{n23}  T_{eff} = \left( \frac{\gamma_{R}}{6 k_{T_{R}}} \left(\frac{1.2}{\pi^2}\right)\frac{m_{P} T_{R}}{\pi \Lambda^2} \right)^{1/3} T_{0} ~.\ee
This again depends only on the ratio $T_{R}/\Lambda^2$. 
For the case of $2$ keV dark matter we find $T_{eff} = 0.30 T_{0}$.

\begin{figure}[htbp]
\begin{center}
\epsfig{file=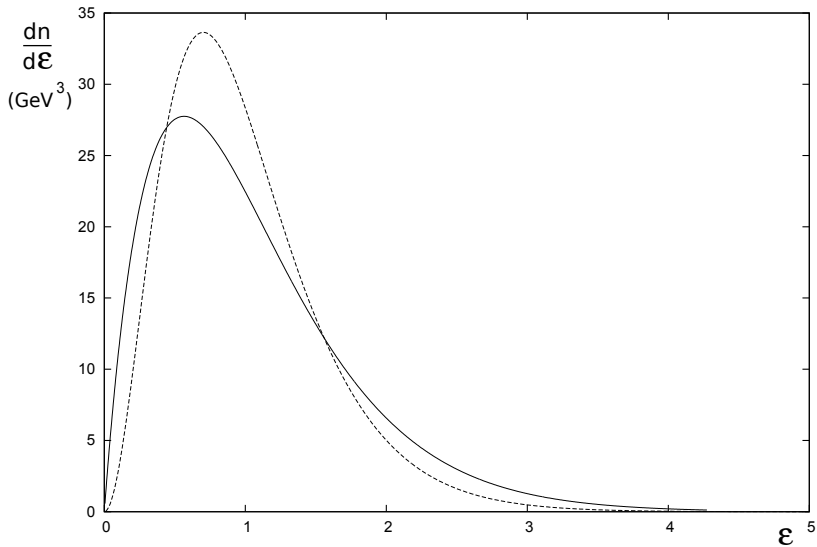, width=0.4\textwidth, angle = -90}
\caption{Comparison of the shape of UV freeze-in distribution (solid line) and the thermal relic fermion distribution (dashed line) as a function of $\epsilon = E/\overline{E}$ for each distribution.}
\label{fig2}
\end{center}
\end{figure}

   The structure formation consequences of the distribution will determined by its shape relative to its mean energy. Although the distribution from UV freeze-in is hotter than the thermal distribution, this simply means that the dark matter fermion has to be proportionally more massive to have the same free-streaming properties, so that the allowed range 2-3 keV for the thermal distribution will become 4.9-7.4 keV for fermions from UV freeze-in in the case where the shapes of the thermal distributions are the same. To compare the shapes of the distributions, in Figure 3 we show the two distributions as a function of $\epsilon = E/\overline{E}$, where the average energy is $\overline{E} = 0.95 T$ for the thermal distribution and $\overline{E} = 2.34 T$ for the UV freeze-in distribution. The distributions are similar, with a slightly greater spread of energies about the mean energy for UV freeze-in WDM. Thus we expect the range of dark matter particle mass from UV freeze-in to be 5-7 keV. The lower bound on the reheating temperature from \eq{d20} for $m_{\psi} = 5$ keV is $T_{R} \gae 120$ GeV, with the corresponding lower bound on $\Lambda$ becoming $\Lambda \gae 6.0 \times 10^9$ GeV.

\section{Conclusions} 

  We have considered the UV freeze-in mechanism for keV mass warm dark matter in some detail, focusing on the case of fermionic Higgs portal dark matter. Perhaps the most interesting result is the dark matter energy distribution from UV freeze-in. We find that it is hotter than the equivalent thermal distribution with the same number density and has the form of a Bose-Einstein distribution with a non-thermal normalization, which is a result of its production from thermal Higgs boson annihilation. This energy distribution can serve as the starting point for studies of structure formation due to WDM from UV freeze-in. 
 
   The shapes of the thermal relic and UV freeze-in distributions are similar, with a slightly greater spread around the mean energy for UV freeze-in WDM. Therefore the WDM properties of UV freeze-in fermionic Higgs portal dark matter will be similar to those of thermal relic fermion dark matter, except for the higher mean energy of the UV freeze-in distribution ($\overline{E} = 2.34 T$ compared to $\overline{E} = 0.95 T$ for thermal relic WDM).  Thus we expect the range of mass for which UV freeze-in WDM is consistent with observations to be scaled up from 2-3 keV for thermal relic WDM to 5-7 keV for UV freeze-in WDM.  The corresponding bounds on the reheating temperature and the Higgs portal interaction scale for $m_{\psi} = 5$ keV are $T_{R} \gae 120$ GeV and $\Lambda \gae 6.0 \times 10^{9}$ GeV, with $T_{R} \approx 120$ GeV for the case where the dark matter fermion gains its mass entirely from electroweak symmetry breaking.

   We finally comment on whether fermionic Higgs portal WDM from UV freeze-in can be observationally tested. The fermions are very weakly coupled to the Standard Model, so no direct experimental test of the model is possible. The only way to confirm the model would be to show that WDM has a Bose-Einstein distribution with a non-thermal normalization, and that the WDM particle is fermionic. This might be possible if observations and simulations of structure formation can  determine in detail the initial momentum distribution of the WDM and if fermionic quantum pressure effects at small scales (as suggested in \cite{sanchez3}) can be confirmed.

\end{document}